\documentclass[twocolumn]{revtex4-1}
\usepackage{amsmath}
\usepackage{graphicx} 
\begin{document}
\title{The space writhes and signatures of polymer knots}
\author{Finn Thompson}
\affiliation{School of Mathematics and Physics, University of Queensland}
\author{Maria Maalouf}
\affiliation{Department of Mathematical Sciences, Smith College}
\author{Alexander R. Klotz}
\affiliation{Department of Physics and Astronomy, California State University, Long Beach}


\begin{abstract}
The space writhe of a knot is a property of its three-dimensional embedding that contains information about its underlying topology, but the correspondence between space writhe and other topological invariants is not fully understood. We perform Langevin dynamics simulations of knotted semiflexible polymers and measure their ensemble average space writhe. We show that for all knots up to 10 crossings, alternating and non-alternating, the average space writhe is almost equal to that of the tightest known configuration of the same knot, with minor differences. Using this equivalence, we show that for more complex knots with up to 38 crossings, the average space writhe is strongly correlated with the signature of the knot. This establishes that the connection between signature and space writhe holds at larger crossing numbers.
\end{abstract}

\maketitle

\section{Introduction}
The space writhe of a tight alternating knot lies near an integer multiple of 4/7 \cite{pieranski2001quasi}, a value that can be predicted from operations on a minimal diagram of the knot \cite{cerf2000topological}. It was recently found that a similar property holds for non-alternating knots, and that the space writhe of a tight knot has a strong but unexplained correlation with the invariant signature of the knot \cite{klotz2024ropelength}. To better understand these features of a knot's space writhe, this paper has three goals. The first is to establish whether the ensemble average writhe of a simulated knotted polymer is equivalent to its tight value (beyond the simple knots already tested \cite{huang2001crossings}), particularly for non-alternating knots. The second is to examine in greater depth the equivalence or lack thereof between the predicted, ideal, and ensemble average writhe of a knot. The third is to verify whether the signature-writhe relationship observed in 12-crossing knots applies at higher crossing number, having established that the polymer writhe can be used for such a measurement.

Knots have invariant properties that do not depend on the specific configuration of the knot. Many of these properties are computed from two-dimensional diagrams of a knot projected onto a surface, rather than the three-dimensional curve of the knot. There is likely additional information contained within a knot's space curve that can also define topological invariants. For example Sleiman et al. \cite{sleiman2023learning} recently showed that a neural network trained on a ``local writhe'' representation of a knotted polymer could determine knot type better than diagram-based invariants. Despite knot invariants being invariant under ambient isotopy, there are several unique 3D configurations that have invariants associated with them. These include the ``ideal" configuration, which minimizes the contour of the knot given a no-overlap constraint, leading to an invariant known as ropelength \cite{ashton2011knot}. There are also unique families of configurations that minimize certain knot energies such as the electrostatic potential or the Mobius energy \cite{hoidn2002quantization}. 

A parameter associated with a 3D configuration of a knot is the space writhe. The writhe of a diagram is the difference between the number of positive and negative crossings. The space writhe is the average over every possible projection of a knot onto all surfaces (Fig. 1). As the Gauss linking integral calculates the net writhe of one curve around another, the space writhe may be thought of as the Gauss linking integral of a curve with itself, or of two identical curves separated by zero distance. Hereafter, the space writhe will just be referred to as the writhe. The writhe of an ideal knot is not minimal, but it is known to lie close to multiples of 4/7 for alternating knots \cite{pieranski2001quasi} and 4/3 for non-alternating knots \cite{klotz2024ropelength}, a property known as quasi-quantization. Cerf and Stasiak \cite{cerf2000topological} showed that a crossing-removal operation on alternating diagrams yielded a ``predicted writhe'' for each knot. They argued that writhe should be quantized not only for ideal knots, but for an ensemble average of many configurations of a knot. 

\begin{figure}
    \centering
    \includegraphics[width=0.8\linewidth]{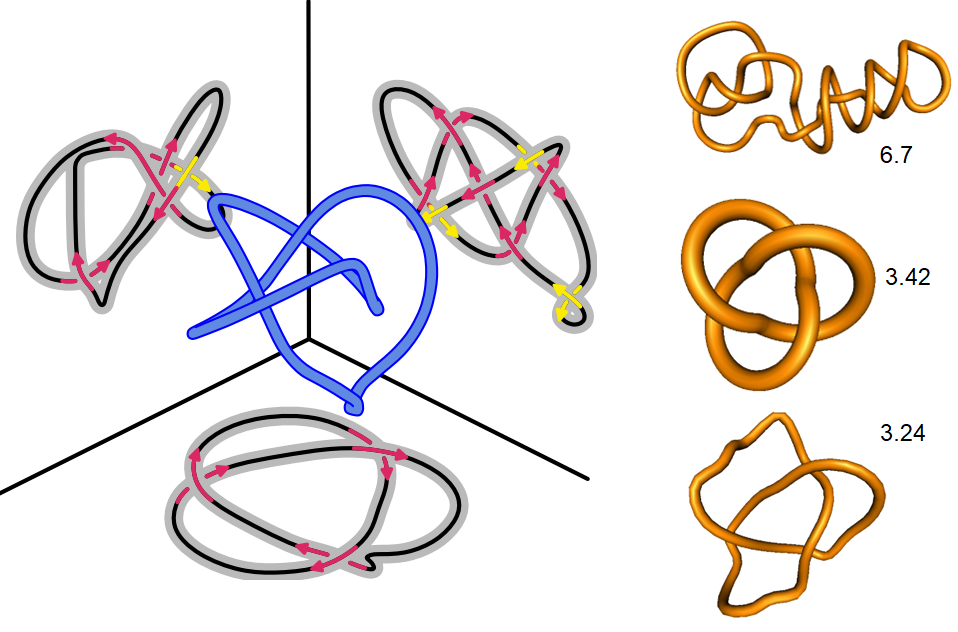}
    \caption{Left: A trefoil knot in three-dimensional space projected onto three perpendicular axes. Each projection has a different number of crossings, which may be positive (red) or negative (yellow). The difference between positive and negative crossings is the writhe of the projected diagram, and the space writhe is the average over all projections. Right: 
    Three projections of a trefoil knot, each with different space writhe: a twisty one with space writhe 6.7, a tight one with a value of 3.42 (close to the predicted 24/7), and a polymer-like one with a value of 3.24.}
    \label{fig:fig1}
\end{figure}

The ensemble average of configurations of a single knot is difficult to define mathematically; an infinite family of isotopies can be constructed by arbitrarily twisting a knot until it has whatever writhe which is desired. In contrast, a polymer chain typically fluctuates through a space of configurations in a manner that maximizes entropy while subject to the constraints of bending rigidity and self-avoidance. A closed polymer chain can be simulated with forces that prevent strands from passing through each other, preserving the initial topology without the use of knot invariants. Polymer chains can be simulated quickly and efficiently and their writhe can be computed from their Cartesian coordinates. They are a useful system for generating an ensemble of knotted isotopies, and pathological configurations (such as those with very sharp bends or extremely close approaches of the curve) are prevented by the physics of the simulation. They have been used to make inferences about the mathematical properties of knots since the simulations became viable around 2000 \cite{huang2001crossings}, and are still used so today \cite{sleiman2023learning}. 
In comparison to polymer knots which must be studied as an ensemble, ropelength-minimizing ideal knots represent a privileged configuration from which properties intrinsic to the knot may be computed. The algorithms to generate them such as SONO \cite{pieranski1998search} and Ridgerunner \cite{ashton2011knot}, however, are considerably slower than polymer simulations and are subject to becoming trapped in local minima. Their computation time also increases with vertex and crossing number more significantly than polymer simulations. For computing the writhe of a knot, the polymer configurations that can be averaged after a few seconds of simulation on a laptop show similar values to the writhes of ideal knots that can take hours to arrive at. There is limited evidence that the ensemble average of simulated polymer knots has the same or a similar writhe as the ideal configuration, but this has only been tested for a few classes of highly symmetric alternating knots such as $(p,2)$ torus knots and twist knots \cite{huang2001crossings, hoidn2002quantization}. We hope to put this equivalence on firmer ground.


The writhe of non-alternating knots is less understood, and initially the idea of writhe quantization was only discussed for alternating knots. Non-alternating knots admit multiple minimal diagrams, making it difficult to predict their writhe based on the crossing nullification procedure. Stasiak showed that averaging the two predicted writhes of the $8_{20}$ knot would predict a writhe of 2 \cite{stasiak2000knots}, which has been observed numerically as well. Recently, Klotz and Anderson \cite{klotz2024ropelength} showed that the writhe of non-alternating knots is distributed around integer or half-integer multiples of 4/3 (rather than 4/7), depending on the parity of the crossing number. A wider distribution about the quantized value was observed than for alternating knots. Here, we attempt to determine whether the ensemble average writhe of non-alternating polymer knots also has the ideal value.

Another finding of Klotz and Anderson \cite{klotz2024ropelength} showed that the ideal writhe is strongly correlated with the signature of the knot, an invariant based on the Seifert surface of the knot, as defined by Trotter \cite{writheDefnTrotter} and Murasugi \cite{writheDefnMurasugi}. For most knots investigated, the signature takes the same value as the Rasmussen s invariant from Khovanov homology \cite{dasbach2011turaev} and it is not known which correlation is more fundamental. The correlation with writhe was over 0.95 in the 2188 knot sample, and suggests a deep connection between the two. This is particularly striking as the variation and correlation are within a single crossing number, negating possible spurious correlations due to crossing number effects. However, trends seen in knots at small crossing numbers do not necessarily hold at higher crossing number. After we establish that the average writhe of a knotted polymer simulation can reproduce the ideal value, we will examine this correlation in greater depth using more complex knots that go beyond the 12-crossing knots examined previously.


\section{Methods}

Knotted polymers were simulated using molecular dynamics by initializing Kremer-Grest bead-spring chains \cite{kremer1990dynamics} with a given knot type and iterating their dynamics by solving the Langevin equation in \textit{LAMMPS} \cite{thompson2022lammps}. Lennard-Jones excluded volume interactions and finitely-extenstible springs preserved the topology by preventing strand passage. The persistence length was set to ten times the bead diameter, to ensure a smooth normal vector along the chain and for consistency with other recent work. Further details are provided in the appendix. Cartesian coordinates of all 249 knots with up to 10 crossings were taken from KnotInfo \cite{knotinfo} and converted to initial coordinate files for the simulations, using spline interpolation to set a common number of beads for each knot. Coordinates were saved at long enough time intervals such that the radius of gyration was uncorrelated between snapshots. For initial investigations, typically 21 uncorrelated snapshots of a 100-bead chain were used to measure the space writhe. Our initial goal was not to generate the most precise possible measurement, but to examine consistency with the ideal writhe, and we typically increase the number of configurations tenfold for more precise measurements. The simulations contained a short equilibration stage at the beginning using Hookean springs, which are more susceptible to strand passage. To check for changes in the knot topology due to pathological strand crossings, we compute the Alexander polynomial at -1 for the final coordinates of the simulation. In cases where it changed, we reran the simulation with more vertices.

We compute the space writhe of each configuration based on the integral formulation:

\begin{equation}
    Wr=\frac{1}{4\pi}\int_{K_1}\int_{K_2}\vec{dr_{1}}\times\vec{dr_{2}}\cdot\frac{\vec{r_{1}}-\vec{r_{2}}}{|\vec{r_{1}}-\vec{r_{2}} |^3},
\end{equation}
where the integral is computed along two identical copies of the knot parameterized in Cartesian space as $r_1$ and $r_2$. Although this can be computed discretely using the tangent vector between adjacent beads, we use the method of solid angles \cite{klenin2000computation} as it is less sensitive to discretization errors:
\begin{equation}{Wr}=\sum_{i=1}^{N}\sum_{j=1}^{N}\frac{\Omega_{ij}}{4\pi}
\end{equation}
where i and j are indices of the line segments that make up the knot. The two line segments define a quadrilateral, with vertices labeled 1 through 4. The solid angle subtended by each quadrilateral is:

\begin{equation}
\begin{alignedat}{1}
\Omega^{*}&=\arcsin\left(n_{1}\cdot n_{2}\right)+\arcsin\left(n_{2}\cdot n_{3}\right) \\
&\quad+\arcsin\left(n_{3}\cdot n_{4}\right)+\arcsin\left(n_{4}\cdot n_{1}\right),
\end{alignedat}
\end{equation}

where

\begin{equation}
\begin{alignedat}{2}
  n_{1}&=\frac{r_{13}\times r_{14}}{\left|r_{13}\times r_{14}\right|},\quad n_{2}&=\frac{r_{14}\times r_{24}}{\left|r_{14}\times r_{24}\right|}, \\
  n_{3}&=\frac{r_{24}\times r_{23}}{\left|r_{24}\times r_{23}\right|},\quad n_{4}&=\frac{r_{23}\times r_{13}}{\left|r_{23}\times r_{13}\right|}
\end{alignedat}
\end{equation}

and the sign of the solid angle is determined by the triple product of the three vectors:
\begin{equation}
    \frac{\Omega}{4\pi}=\frac{\Omega^{*}}{4\pi}\text{sign}\left(\left(r_{34}\times r_{12}\right)\cdot r_{13}\right).
\end{equation}
The sign of the writhe is reversed by reflection, and the sign of the average writhe of a chiral knot depends on which stereoisomer is examined. The writhes of an ensemble may be distributed on both sides of zero. We take the absolute value of the average of measured writhes to choose a single chirality. Ideal writhes were measured based on the tight configurations provided in Fourier form by Gilbert on the Knot Atlas wiki \cite{katlas}. Outlying values were compared to writhes calculated from the coordinates of Ashton et al. \cite{ashton2011knot}

To explore knots more complex than those currently tabulated, we used \textit{SnapPy} \cite{snappy} to generate a series of random single component links with crossing numbers ranging from 13 to 38. By default, \textit{SnapPy} applies a basic level of simplification to each knot, so these crossing numbers are close to the minimum value for each knot. Additionally, the signatures and Dowker-Thistlethwaite (DT) codes were generated for each knot.
Using \textit{KnotPlot} \cite{scharein2002interactive}, each DT code was used to generate simple 3D coordinates representing each knot. For a knot with $n$ crossings, \textit{KnotPlot} generates an embedding of the knot defined by $8n$ points. However, this embedding is frequently not smooth, and the distance between non-adjacent lines connecting pairs of points can be arbitrarily small. Hence, \textit{KnotPlot}'s built-in dynamic physics engine was used to smooth each knot. This procedure acts as a discrete ambient isotopy, preserving the type of each knot. In particular, a low ``mech'' (attractive force between adjacent points), high ``elec'' (repulsive force between non-adjacent points) and a varying ``therm'' (random agitation of points) were used, evolving the knot until its diagram appeared `smooth'. The coordinates of the resulting knots were saved.
Each of these smoothed knots still have a crossing number that is close to its minimum crossing number, and typically 200-300 vertices. The knots were simulated as semi-flexible polymers in \textit{LAMMPS}, to generate a sample of configurations of each knot. Using these methods, we measured the average space writhe of 74 alternating knots and 23 non-alternating knots. The complexity of the knots that can be generated by this procedure is limited by \textit{KnotPlot}'s DT input algorithm. \textit{SnapPy} can also generate non-alternating diagrams. However, non-alternating diagrams may be non-minimal projections of alternating knots. In such cases, we may use \textit{SnapPy}’s in-built simplification methods which use random Reidemeister moves to determine if the diagram reduces to one of a smaller alternating knot. We have investigated the subset of non-alternating diagrams that remain non-alternating after simplification.


\section{Results and Discussion}

\begin{figure}
    \centering
    \includegraphics[width=1\linewidth]{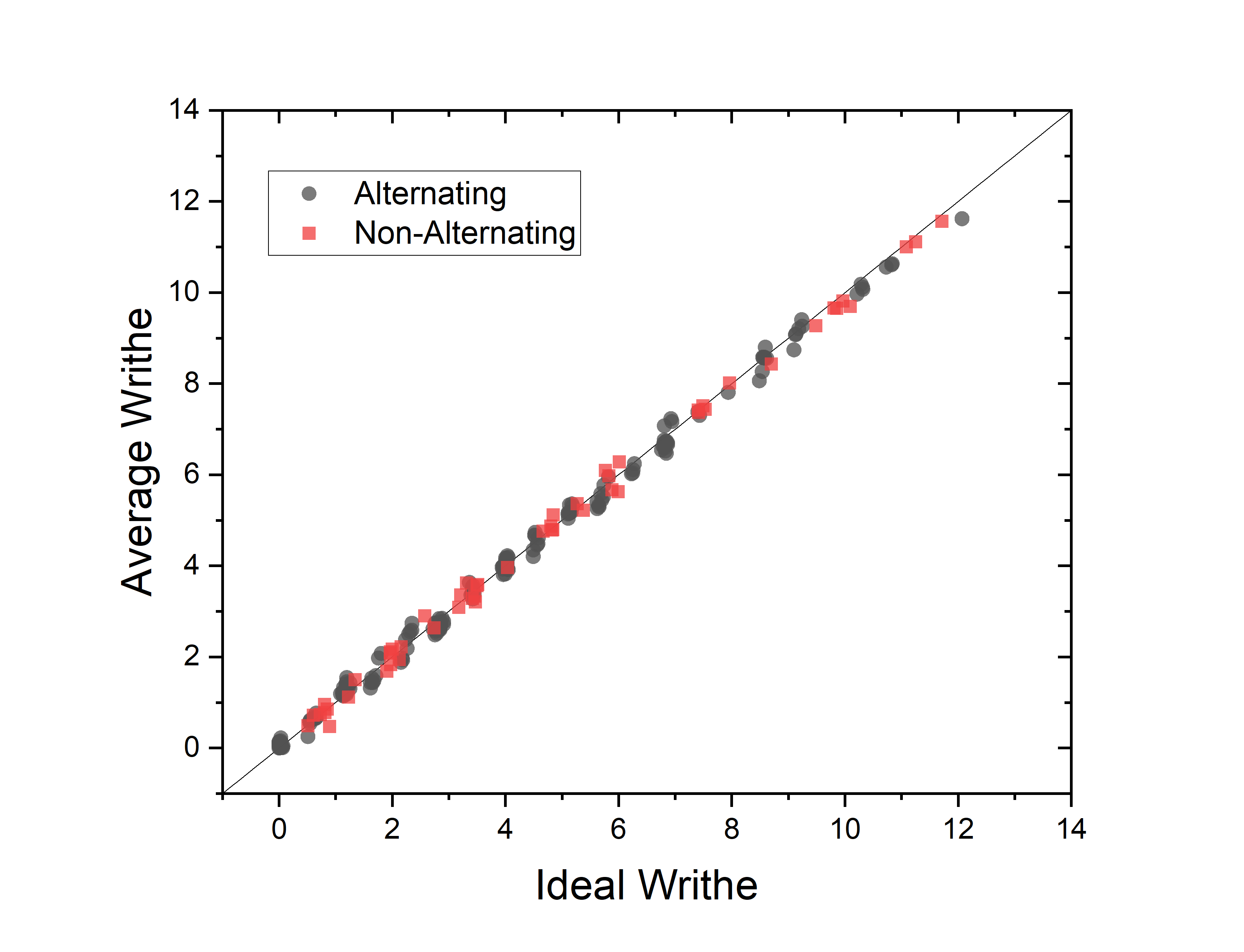}
    \caption{Correlation between the ideal writhe of a knot and the average writhe of a polymer knot for alternating (black) and non-alternating (red) knots. The line represents equality.}
    \label{fig:polymervsideal}
\end{figure}

Figure \ref{fig:polymervsideal} shows the correlation between the average writhe of a polymer knot and the writhe of the ideal configuration for all knots between 3 and 10 crossings. The correlation is quite strong (0.999), with minor deviations about the line of equality. While this has been established for some simplified families of knots \cite{huang2001crossings}, we have now established this correlation for non-alternating knots and for knots that do not belong to a highly symmetric family such as alternating torus knots. Why does the ensemble average writhe of a knotted polymer converge on the ideal value? An argument from Yuanan Diao (personal communication) is that there exists some fundamental configuration of a knot with a certain writhe (be it the ideal or predicted value), and families of operations that can perturb the knot and change the writhe. However, for every operation that can increase the writhe, there is an opposite operation that will decrease it by the same amount. The average after every possible operation on the knot is simply that fundamental value. Although the data in Fig. 2 is not particularly precise, the writhes of the alternating polymer knots are biased towards multiples of 4/7, and the writhes of the even-crossing non-alternating polymer knots are biased towards half-integer multiples of 4/3, suggesting that writhe quantization applies to polymer knots as well. As mentioned, it is generally much quicker computationally to perform a Langevin dynamics simulation of a polymer than it is to perform constrained gradient optimization on the knot, and we have established that the former is a valid method to reproduce the results of the latter.

\begin{figure}
    \centering
    \includegraphics[width=1\linewidth]{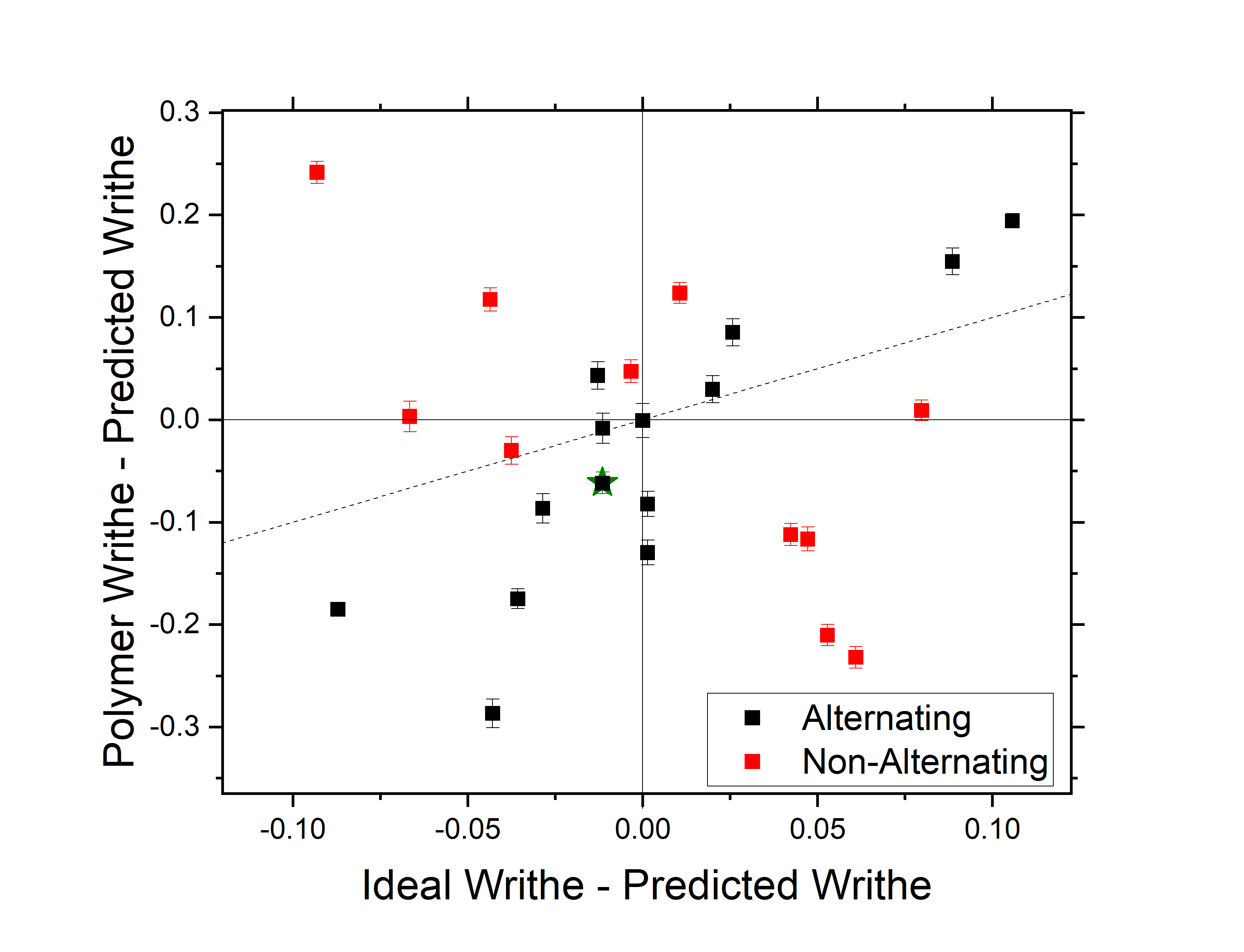}
    \caption{Scatter plot of the deviation between the ideal and predicted writhe on the x-axis and between the polymer and predicted writhe on the y-axis, for alternating knots up to 7 crossings (black) and non-alteranting knots with 8 and 9 crossings (red). The diagonal line is y=x. The green star is from a sample of random stick trefoil knots.}
    \label{fig:deviation}
\end{figure}

The ideal writhe is known to slightly deviate from the predicted writhe of Cerf and Stasiak \cite{cerf2000topological}, even for very tight knots (the tightest known trefoil \cite{przybyl2014high} has a writhe of 23.92/7, compared to 24/7). Does the average polymer writhe converge to the predicted writhe, or the ideal writhe? Averaging many statistically independent configurations will yield a mean value with a standard error. With 200 configurations averaged, the standard error is typically 0.01 to 0.02, much smaller than the 0.57 between successive alternating writhe quanta, but comparable to or slightly smaller than observed differences between the predicted and ideal writhe. Examining knots with greater deviations between ideal and predicted writhes, even after 2000 averaged configurations, the polymer writhe does not appear to converge to either value. For the $6_2$ knot the values of the predicted, ideal, and polymer writhe are 2.857, 2.785, and 2.672$\pm$0.006, and for the $7_4$ knot they are 5.714, 5.784, and 5.909$\pm$0.007. Fig. \ref{fig:deviation} shows a higher precision comparison of predicted, ideal, and polymer writhe. Knots are scattered according to the difference between their ideal writhe and the closest multiple of 4/7 (or 4/3) and the difference between the average polymer writhe and the same. The horizontal line indicates knots for which the ideal form has the predicted writhe, the vertical line indicates knots for which the polymer writhe has the predicted value, and the diagonal line indicates knots for which the ideal and polymer writhes have the same value. The knot for which all writhes are the same is $4_1$. From this data, neither the ideal nor polymer writhe typically have exactly the predicted value, nor the same value, but the deviations are correlated. Interestingly, the correlation is positive for alternating knots and negative for non-alternating knots.

It may be argued that increasing the number of beads in the simulated polymer will improve precision or accuracy. The simulated polymers likely display a minimum crossing number in projection that is not significantly greater than the minimum for each knot, and more may be learned by studying complex configurations with many more crossings than the minimum. However, an interesting feature of polymer knots is self-tightening \cite{grosberg2007metastable}, in which the knot localizes in a small subset of a chain to increase the entropy of the rest of the chain, such that the modal knot size is independent of chain length. Larger chains will not necessarily provide more information about the knots in them. 

These results must be discussed in the context of model dependence. It would be unreasonable to claim that Kremer-Grest chains of length 100 and persistence length 10 are uniquely suited to the study of knots, but it would also be unreasonable to claim that the results cannot be generalized. As an example, we can choose an ensemble of random knots with very little in common with semiflexible polymer knots, and observe the same results. Knots can be stochastically generated by connecting points on the surface of a sphere with line segments, and randomly generated points on a sphere have a chance of generating random knots that represent very different embeddings than polymers. We generated random polygonal knots by choosing 15 random Gaussian unit vectors, corresponding to 15 uniformly distributed points on the unit sphere, and connecting them as line segments. We computed the Alexander polynomial at -1 and +2 and selected the stick-spheres consistent with certain knots. The space writhe of the population of 17761 random sphere trefoils was broadly distributed, but the ensemble average of its absolute value was 3.38$\pm$0.01 (green star in Fig. \ref{fig:deviation}), similar to the polymer value. For 199 $5_1$ stick-sphere knots, the measured average writhe was 6.27$\pm$.08 and the expected value is 6.29. A similar result was reported for spherically-confined random walks by Diao et al. \cite{diao2018average}. This supports the claim that the ensemble average writhe, of a given ensemble, will be close to the predicted writhe (deviating by a small fraction of 4/7), but we do not yet have the information to know how those deviations depend on the model.

Numerical evidence indicates that neither polymer knots nor ideal knots converge exactly on the predicted writhe. Is there a knot type that does? We have not investigated Mobius energy minimizing knots in depth, but measurements by Hoidn et al. \cite{hoidn2002quantization} showed that the writhe of Mobius-minimizing alternating torus knots increased with crossing number as 1.2C, and the predicted writhe increases as $(10/7)\approx1.43C$, so Mobius knots do not have the predicted writhe. Interestingly, every measurement of the $4_1$ knot's writhe, whether it be ideal, polymer, or Mobius, shows consistently zero writhe. This is also observed for the highly symmetric $8_{18}$ knot, but measurements of other amphichiral knots have not been as precisely close to zero.
\begin{figure}
    \centering
    \includegraphics[width=1\linewidth]{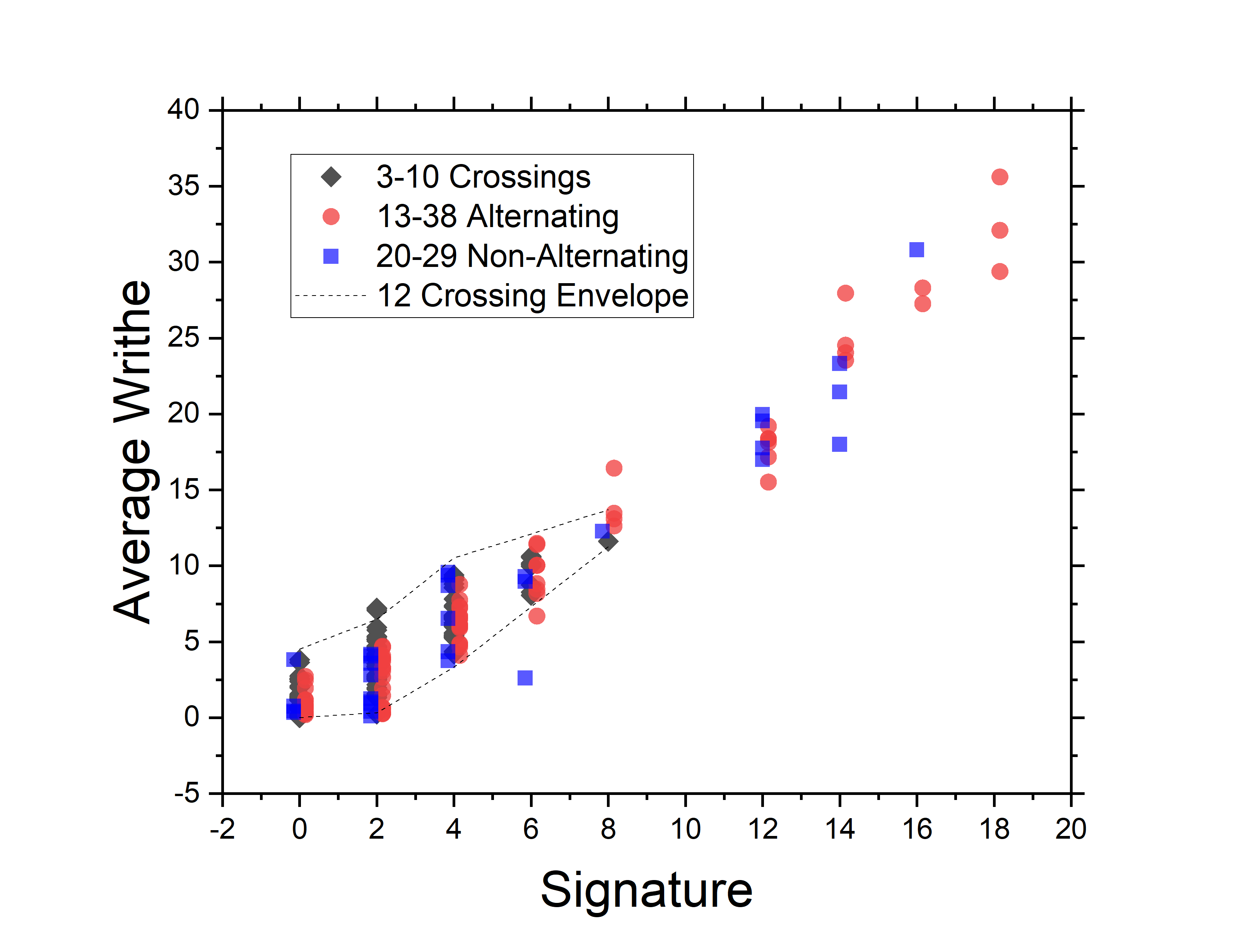}
    \caption{Correlation between the signature of a knot and its average polymer writhe. Black points represent every knot between 3 and 10 crossings, red points represent selected alternating knots with 13-38 crossings, and blue points represent non-alternating knots with 20-29 crossings. The envelope from 1288 12-crossing knots \cite{klotz2024ropelength} is shown for reference. Each set is shifted horizontally by a small amount for visual clarity.}
    \label{fig:siggy}
\end{figure}

Ultimately, we wish to be able to use polymer knots to learn about topological features of knots in general, such as the correlations between invariants. Having established that the polymer writhe is almost the same as the ideal writhe, we can now measure it for more complex knots and examine the signature-writhe correlation at higher crossing number.  Figure \ref{fig:siggy} shows the correlation between the average polymer space writhe and the absolute signature of the knots. Similar to our use of the absolute average writhe, we use the absolute of value of signature and Rasmussen s. The complete sample of knots up to 10 crossings, as well as randomly generated alternating knots up to 38 crossings and non-alternating knots up to 29 crossings, generally follow the same trend, consistent with what was observed for ideal twelve-crossing knots. The absolute signature is bound by the crossing number \cite{ishihara2024signature}, although no 12-crossing knot has an absolute signature greater than $\pm 8$. More complex knots with comparable signatures largely have writhes that are within the bounds observed in 3-12 crossing knots. The exceptions do not appear to be statistical fluctuations and persist when more configurations are averaged, and likely indicate a bigger spread in the writhes of more complex knots. There is no apparent difference in the relation between alternating and non-alternating knots, just as there was not for 12-crossing knots. Alternating knots with higher signatures have writhes that follow the same trend, with a spread that is again greater than the simpler knots. The overall correlation between signature and writhe for all the complex knots was 0.96. We take this as evidence that the correlation between writhe and signature persists for more complex knots.

For non-alternating knots, the signature is not necessarily equal to the Rasmussen s invariant. Among the 105 12-crossing knots for which signature and s are different, the ideal writhe has a 0.98 correlation with s and a 0.85 correlation with signature. Using the software \textit{KnotJob} \cite{schutz2022scanning} we computed s for our complex non-alternating knots, and found five cases where the two invariants differed. We did not have sufficient data to investigate this for our complex knots: the correlation was slightly stronger with signature, but the change in correlation was effectively due to a single knot with a large writhe having a signature of 16 and an s of 12. It is also difficult to disentangle the effect of crossing number, so we cannot state with certainty at this time which invariant is the more fundamentally related to writhe. 

We still do not have a full explanation for the writhe-signature correlation. Previously, we speculated that a set of embeddings exist that minimize the area of a knot's Seifert surface (given suitable constraints), that these configurations are dictated by the genus of the surface which also dictates the signature, and that such an embedding would have a writhe comparable to the ideal one. We hope this work inspires experts to uncover a more complete explanation.

It is an open question whether the writhe quantization phenomenon holds for more complex knots. The more complex knots we have examined do not display this feature. This may be due to the weakening of the quantization with knot complexity, or simply the difficulty of obtaining precise measurements for larger knots, but our statistical uncertainty in the writhe measurement is considerably less than 4/7 or 4/3.

\section{Conclusion}
We have investigated the writhe of polymer knots using Langevin dynamics simulations. We sought to verify that the average polymer writhe is equivalent to the ideal writhe for more complex and non-alternating knots, and did so, showing a 0.999 correlation. We sought to investigate whether the average writhe of many configurations converges closer to the predicted value than the ideal writhe does and found evidence against that, showing that the average polymer writhe deviates from the predicted writhe in a similar manner as the ideal configuration. Finally, we sought to extend evidence for the writhe-signature correlation to more complex knots at higher crossing number, and verified that the same relation observed in knots up to 12 crossings holds for alternating and non-alternating knots up to 38 crossings. Having put on firmer ground the use of polymer simulations to generate meaningful knot embeddings, we hope such methods can be used to investigate other relations between invariants and 3-space coordinates in the future. Additionally, we hope that these results encourage researchers to uncover a more fundamental explanation for this correlation.

\section{Acknowledgements}
ARK is supported by the National Science Foundation, award no. 2336744.

\bibliographystyle{unsrt}
\bibliography{writherefs}

\begin{thebibliography}{10}

\bibitem{pieranski2001quasi}
Piotr Piera{\'n}ski and Sylwester Przyby{\l}.
\newblock Quasi-quantization of writhe in ideal knots.
\newblock {\em The European Physical Journal E}, 6(2):117--121, 2001.

\bibitem{cerf2000topological}
Corinne Cerf and Andrzej Stasiak.
\newblock A topological invariant to predict the three-dimensional writhe of ideal configurations of knots and links.
\newblock {\em Proceedings of the National Academy of Sciences}, 97(8):3795--3798, 2000.

\bibitem{klotz2024ropelength}
Alexander~R Klotz and Caleb~J Anderson.
\newblock Ropelength and writhe quantization of 12-crossing knots.
\newblock {\em Experimental Mathematics}, pages 1--8, 2024.

\bibitem{huang2001crossings}
Juin-Yan Huang and Pik-Yin Lai.
\newblock Crossings and writhe of flexible and ideal knots.
\newblock {\em Physical Review E}, 63(2):021506, 2001.

\bibitem{sleiman2023learning}
Joseph~Lahoud Sleiman, Filippo Conforto, Yair Augusto~Gutierrez Fosado, and Davide Michieletto.
\newblock Geometric learning of knot topology.
\newblock {\em Soft Matter}, 20(1):71--78, 2024.

\bibitem{ashton2011knot}
Ted Ashton, Jason Cantarella, Michael Piatek, and Eric~J Rawdon.
\newblock Knot tightening by constrained gradient descent.
\newblock {\em Experimental Mathematics}, 20(1):57--90, 2011.

\bibitem{hoidn2002quantization}
Phoebe Hoidn, Robert~B Kusner, and Andrzej Stasiak.
\newblock Quantization of energy and writhe in self-repelling knots.
\newblock {\em New Journal of Physics}, 4(1):20, 2002.

\bibitem{pieranski1998search}
Piotr Piera{\'n}ski.
\newblock In search of ideal knots.
\newblock {\em Computational Methods in Science and Technology}, 4:9--23, 1998.

\bibitem{stasiak2000knots}
A~Stasiak.
\newblock Knots in hellas’ 98, 2000.

\bibitem{writheDefnTrotter}
H.~F. Trotter.
\newblock Homology of group systems with applications to knot theory.
\newblock {\em Ann. of Math. (2)}, 76:464--498, 1962.

\bibitem{writheDefnMurasugi}
Kunio Murasugi.
\newblock On a certain numerical invariant of link types.
\newblock {\em Trans. Amer. Math. Soc.}, 117:387--422, 1965.

\bibitem{dasbach2011turaev}
Oliver~T Dasbach and Adam~M Lowrance.
\newblock Turaev genus, knot signature, and the knot homology concordance invariants.
\newblock {\em Proceedings of the American Mathematical Society}, pages 2631--2645, 2011.

\bibitem{kremer1990dynamics}
Kurt Kremer and Gary~S Grest.
\newblock Dynamics of entangled linear polymer melts: A molecular-dynamics simulation.
\newblock {\em The Journal of Chemical Physics}, 92(8):5057--5086, 1990.

\bibitem{thompson2022lammps}
Aidan~P Thompson, H~Metin Aktulga, Richard Berger, Dan~S Bolintineanu, W~Michael Brown, Paul~S Crozier, Pieter~J In't~Veld, Axel Kohlmeyer, Stan~G Moore, Trung~Dac Nguyen, et~al.
\newblock Lammps-a flexible simulation tool for particle-based materials modeling at the atomic, meso, and continuum scales.
\newblock {\em Computer Physics Communications}, 271:108171, 2022.

\bibitem{knotinfo}
Charles Livingston and Allison~H. Moore.
\newblock Knotinfo: Table of knot invariants.
\newblock URL: \url{knotinfo.math.indiana.edu}, May 2023.

\bibitem{klenin2000computation}
Konstantin Klenin and J{\"o}rg Langowski.
\newblock Computation of writhe in modeling of supercoiled dna.
\newblock {\em Biopolymers: Original Research on Biomolecules}, 54(5):307--317, 2000.

\bibitem{katlas}
Dror Bar-Natan and Brian Gilbert.
\newblock Knot atlas. ideal knots. www.katlas.org/wiki/ideal\_knots.

\bibitem{snappy}
Marc Culler, Nathan~M. Dunfield, Matthias Goerner, and Jeffrey~R. Weeks.
\newblock Snap{P}y, a computer program for studying the geometry and topology of $3$-manifolds.
\newblock Available at \url{http://snappy.computop.org} (15/12/2024).

\bibitem{scharein2002interactive}
Robert~G Scharein and Kellogg~S Booth.
\newblock Interactive knot theory with knotplot.
\newblock In {\em Multimedia Tools for Communicating Mathematics}, pages 277--290. Springer, 2002.

\bibitem{przybyl2014high}
Sylwester Przybyl and Piotr Pieranski.
\newblock High resolution portrait of the ideal trefoil knot.
\newblock {\em Journal of Physics A: Mathematical and Theoretical}, 47(28):285201, 2014.

\bibitem{grosberg2007metastable}
Alexander~Y Grosberg and Yitzhak Rabin.
\newblock Metastable tight knots in a wormlike polymer.
\newblock {\em Physical review letters}, 99(21):217801, 2007.

\bibitem{diao2018average}
Yuanan Diao, Claus Ernst, Eric~J Rawdon, and Uta Ziegler.
\newblock Average crossing number and writhe of knotted random polygons in confinement.
\newblock {\em Reactive and Functional Polymers}, 131:430--444, 2018.

\bibitem{ishihara2024signature}
Kai Ishihara, Kei Okada, and Koya Shimokawa.
\newblock Signature and crossing number of links.
\newblock {\em arXiv preprint arXiv:2410.00445}, 2024.

\bibitem{schutz2022scanning}
Dirk Sch{\"u}tz.
\newblock A scanning algorithm for odd khovanov homology.
\newblock {\em Algebraic \& Geometric Topology}, 22(3):1287--1324, 2022.

\end{thebibliography}

\section{Appendix}

We simulate polymer knots with $N$ monomers using a commonly-used model, and our descriptions bear similarity to previous descriptions of these methods including our own. Polymers are comprised of beads of diameter $\sigma$ (which sets the lengthscale of the system) at position $\mathbf{r}_i(t)$, connected by springs to their two neighbors. A finitely-extensible nonlinear elastic (FENE) spring potential with a maximum extension of $1.5 \sigma$ is used. Excluded volume interactions between beads are enforced by a truncated Lennard-Jones repulsive potential that applies when the centers of mass of two beads are closer than 1.122$\sigma$. The relatively short range of distances between the excluded volume of the beads and maximum extension of the springs ensures that strands do not cross and the link topology is preserved. Bending rigidity is imposed by a Kratky-Porod potential depending on the cosine of the angle between three successive beads. The strength of this potential sets the persistence length of the polymer. The entire contribution to the energy of a bead is:
\begin{equation}
    U_{\rm tot}=U_{\rm spr}+U_{\rm ev}+U_{\rm bend}\, .
\end{equation}
The excluded volume interaction takes the form:
\begin{equation}
    U_{\rm ev}=\begin{cases}
  4\epsilon\left[\left(\frac{\sigma}{r}\right)^{12}-\left(\frac{\sigma}{r}\right)^{6}+\frac{1}{4}\right]
    & \text{if } r\leq 2^{1/6}\sigma\\
    0              & \text{otherwise},
\end{cases}
\end{equation}
where $\epsilon$ sets the energy scale of the repulsive interactions. The spring force is parameterized as:
\begin{equation}
    U_{\rm spr}=\begin{cases}
    -\frac{1}{2}\left(\kappa\frac{\epsilon}{\sigma^2}\right)R_{\rm max}\log\left| 1-\left(\frac{r}{R_{\rm max}}\right)^{2}\right|,  & \text{if } r\leq R_{\rm max}\\
    \infty             & \text{otherwise},
\end{cases}
\end{equation}
where $\kappa$ is 30 and sets the spring constant in units of $\epsilon/\sigma^2$, and $R_{\rm max}=1.5\sigma$ is the maximum separation of the springs. The bending potential takes the form:
\begin{equation}
    U_{\rm bend}=\frac{\ell_{p}}{\sigma}kT(1-\cos{\theta})\,.
\end{equation}
The dimensionless ratio of the persistence length $\ell_{p}$ to the bead diameter is $\ell_p/\sigma = 10$ in this work. The time evolution of the $i$\textsuperscript{th} bead is determined by the Langevin equation:
\begin{equation}
    m\ddot{\mathbf{r}}_i(t) =-\gamma\dot{\mathbf{r}}_i(t) - \nabla_{\mathbf{r}_i} U_{\rm tot}+\sqrt{2kT\gamma}\,\bm{\eta}(t)\,.
\end{equation}
Here, $\gamma$ is the drag coefficient on a single bead, $kT$ is the thermal energy scale, $\bm{\eta}$ is a delta-correlated normal random variable, i.e.~$\langle \eta_i(t) \eta_j(t') \rangle = \delta_{ij}\delta(t-t')$, and an overdot represents a time derivative. The final term provides a random force that emulates Brownian motion in a manner consistent with the fluctuation-dissipation theorem. These equations of motion are solved by \textit{LAMMPS}~\cite{thompson2022lammps}, which iterates the system forward in time using the Velocity Verlet algorithm.

The system is non-dimensionalized with $\sigma$, $\gamma$, $m$, $kT$ and $\epsilon$ taking values of 1, which defines a timescale $\tau=\sigma\sqrt{m/\epsilon}$. We iterate the simulation with a timestep of 0.01 $\tau$. We initially perform 200 iterations of the system with a simpler harmonic spring potential to avoid overstretched FENE springs. Then, we typically perform one million iterations (10,000 $\tau_{LJ}$), saving the coordinates every 50,000.

\end{document}